\documentclass[twocolumn,showpacs,preprintnumbers,amsmath,amssymb]{revtex4}
\usepackage{graphicx}  %for pdflatex use .jpg plots

\begin{document} 
\renewcommand{\textfraction}{0.6}
\renewcommand{\dbltopfraction}{0.4}
\renewcommand{\bottomfraction}{0.4}
\renewcommand{\dblfloatpagefraction}{0.4}
\topmargin= -0.1 true in
%\draft 

%%%% code for the organisms
\def\Atha{{\it Arabidopsis thaliana}}  \def\atha{{\it A. thaliana}}
\def\Aaeo{{\it Aquifex aeolicus}}  \def\aaeo{{\it A. aeolicus}}
\def\Aful{{\it Archaeoglobus fulgidus}}    \def\aful{{\it A. fulgidus}}  
\def\Aper{{\it Aeropyrum pernix}}  \def\aper{{\it A. pernix}} 
\def\Atum{{\it Agrobacterium tumefaciens}}  \def\atum{{\it A. tumefaciens}} 
\def\Bbur{{\it Borrelia burgdorferi}}  \def\bbur{{\it B. burgdorferi}}  
\def\Bhal{{\it Bacillus halodurans}} \def\bhal{{\it B. halodurans}} 
\def\Bmel{{\it Brucella melitensis}}   \def\Bmel{{\it B. melitensis}}   
\def\Bsub{{\it Bacillus subtilis}} \def\bsub{{\it B. subtilis}} 
\def\Busp{{\it Buchnera sp. APS}}   \def\busp{{\it B. sp.}}   
\def\Baph{{\it Buchnera aphidicola}}   \def\baph{{\it B. aphidicola}}   
\def\Cele{{\it Caenorhabditis elegans}} \def\cele{{\it C. elegans}}%nematode
\def\Cjej{{\it Campylobacter jejuni}} \def\cjej{{\it C. jejuni}} 
\def\Ccre{{\it Caulobacter crescentus}} \def\ccre{{\it C. crescentus}} 
\def\Cvib{{\it Caulobacter vibrioides}}  \def\cvib{{\it C. vibrioides}} 
\def\Clim{{\it Chlorobium limicola}} \def\clim{{\it Ch. limicola}} 
\def\Cmur{{\it Chlamydia muridarum}}  \def\cmur{{\it C. muridarum}}  
\def\Cace{{\it Clostridium acetobutylicum}}  \def\cace{{\it C. acetobutylicum}}
\def\Cper{{\it Clostridium perfringens}}  \def\cper{{\it C. perfringens}}   
\def\Cpne{{\it Chlamydia pneumoniae}}  \def\cpne{{\it Ch. pneumoniae}}  
\def\Ctra{{\it Chlamydia trachomatis}} \def\ctra{{\it Ch. trachomatis}} 
\def\Cglu{{\it Corynebacterium glutamicum}} \def\cglu{{\it C. glutamicum}}  
\def\Drad{{\it Deinococcus radiopugans}} \def\drad{{\it D. radiopugans}} 
\def\Dmel{{\it Drosophila melanogaster}} \def\dmel{{\it D. melanogaster}} 
\def\Ecol{{\it Escherichia coli}} \def\ecol{{\it E. coli}} 
\def\Fhep{{\it Flavobacterium heparinum}} \def\fhep{{\it F. heparinum}} 
\def\Fnuc{{\it Fusobacterium nucleatum}} \def\fnuc{{\it F. nucleatum}} 
\def\Gmax{{\it Glycine max}}  \def\gmax{{\it G. max}}    % Soybean
\def\Hasp{{\it Halobacterium sp.}}  \def\hasp{{\it H. sp.}} 
\def\Haur{{\it Herpetosiphon aurantiacus}} \def\haur{{\it H. aurantiacus}} 
\def\Hinf{{\it Haemophilus influenzae}} \def\hinf{{\it H. influenzae}} 
\def\Hpyl{{\it Helicbacter pylori}}  \def\hpyl{{\it H. pylori}}  
\def\Hsap{{\it Homo sapiens}} \def\hsap{{\it H. sapiens}} 
\def\Hvol{{\it Halobacterium volcanii}}  \def\hvol{{\it H. volcanii}}  
\def\Llac{{\it Lactococcus lactis}}  \def\llac{{\it L. lactis}}   
\def\Lmon{{\it Listeria monocytogenes}} \def\lmon{{\it L. monocytogenes}} 
\def\Mlot{{\it Mesorhizobium loti}}  \def\mlot{{\it M. loti}}  
\def\Mfer{{\it Methanothermus fervidus}} \def\mfer{{\it M. fervidus}} 
\def\Mthe{{\it Methanobacterium thermoautotrophicum}} 
                     \def\mthe{{\it M. thermoautotrophicum}}  
\def\Mmus{{\it Mus musculus}} \def\mmus{{\it M. musculus}} % House mouse
\def\Mgen{{\it Mycoplasma genitalium}} \def\mgen{{\it M. genitalium}} 
\def\Mpen{{\it Mycoplasma penetrans}} \def\mpen{{\it M. penetrans}} 
\def\Mpne{{\it Mycoplasma pneumoniae}} \def\mpne{{\it M. pneumoniae}} 
\def\Mpul{{\it Mycoplasma pulmonis}} \def\mpul{{\it M. pulmonis}} 
\def\Mlep{{\it Mycobacterium leprae}} \def\mlep{{\it M. leprae}} 
\def\Mtub{{\it Mycobacterium tuberculosis}} \def\mtub{{\it M. tuberculosis}} 
\def\Nmen{{\it Neisseria meningitidis}} \def\nmen{{\it N. meningitidis}} 
\def\Neis{{\it Neisseria}}
\def\Nost{{\it Nostoc sp.}} \def\nost{{\it N. sp.}} % sp. PCC 7120 
\def\Paby{{\it Pyrococcus abyssi}}  \def\paby{{\it P. abyssi}}   
\def\Paero{{\it Pyrobaculum aerophilum}}  \def\paero{{\it P. aerophilum}}   
\def\Paeru{{\it Pseudomonas aeruginosa}} \def\paeru{{\it P. aeruginosa}} 
\def\Pfur{{\it Pyrococcus furiosus}}  \def\pfur{{\it P. furiosus}}   
\def\Phor{{\it Pyrococcus horikoshii}} \def\phor{{\it P. horikoshii}} 
\def\Pmul{{\it Pasteurella multocida}} \def\pmul{{\it P. multocida}} 
\def\Rcon{{\it Rickettsia conorii}} \def\rcon{{\it R. conorii}} 
\def\Rpro{{\it Rickettsia prowazekii}} \def\rpro{{\it R. prowazekii}} 
\def\Rsol{{\it Ralstonia solanacearum}} \def\rsol{{\it R. solanacearum}} 
\def\Saur{{\it Staphylococcus aureus}} \def\saur{{\it S. aureus}} 
\def\Sent{{\it Salmonella enterica}} \def\sent{{\it S. enterica}} 
\def\Scer{{\it Saccharomyces cerevisiae}} 
                     \def\scer{{\it S. cerevisiae}} % c./yeast
\def\Smel{{\it Sinorhizobium meliloti}}  \def\smel{{\it S. meliloti}}  
\def\Smel{{\it Sinorhizobium meliloti}}  \def\smel{{\it S. meliloti}}  
\def\Sfle{{\it Shigella flexneri}} \def\sfle{{\it S. flexneri}}  
\def\Sone{{\it Shewanella oneidensis}} \def\sone{{\it S. oneidensis}}  
\def\Spne{{\it Streptococcus pneumoniae}} \def\spne{{\it S. pneumoniae}}  
\def\Spyo{{\it Streptococcus pyogenes}} \def\spyo{{\it S. pyogenes}} 
\def\Scoe{{\it Streptomyces coelicolor}} \def\scoe{{\it S. coelicolor}} 
\def\Ssol{{\it Sulfolobus solfataricus}} \def\ssol{{\it S. solfataricus}} 
\def\Stok{{\it Sulfolobus tokodaii}} \def\stok{{\it S. tokodaii}}  
\def\Stub{{\it Solanum tuberosum}} \def\stub{{\it S. tuberosum}} % white potato
\def\Styp{{\it Salmonella typhimurium LT2}} \def\styp{{\it S. typhimurium}} 
\def\Sent{{\it Salmonella enterica}} \def\sent{{\it S. enterica}} 
\def\Syne{{\it Synechococcus sp.}} \def\syne{{\it S. sp.}}
\def\Taci{{\it Thermoplasma acidophilum}}  \def\taci{{\it T. acidophilum}}   
\def\Tmar{{\it Thermotoga maritima}} \def\tmar{{\it T. maritima}} 
\def\Tpal{{\it Treponema pallidum}} \def\tpal{{\it T. pallidum}} 
\def\Tten{{\it Thermoprotues tenax}}  \def\tten{{\it T. tenax}}  
\def\Tvol{{\it Thermoplasma volcanium}} \def\tvol{{\it T. volcanium}}  
\def\Telo{{\it Thermosynechococcus elongatus}} \def\telo{{\it T. elongatus}}  
\def\Uure{{\it Ureaplasma urealyticum}} \def\uure{{\it U. urealyticum}} 
\def\Vcho{{\it Vibrio cholerae}} \def\vcho{{\it V. cholerae}} 
\def\Xfas{{\it Xylella fastidiosa}} \def\xfas{{\it X. fastidiosa}} 
\def\Ypes{{\it Yersinia pestis}}  \def\ypes{{\it Y. pestis}}  
\def\ecols{{\it E. col.}}  \def\mjans{{\it M. jan.}} 
\def\cmurs{{\it C. mur.}}  \def\tpals{{\it T.pal}} \def\afuls{{\it A. ful.}}
%%%%%%%%% End of genome names %%%%%%%%%%%%%%

\def\newpar{{\par\noindent}}
\def\skipaline{{\vskip 12pt plus 1pt}}
\def\skiphafline{{\vskip 6pt plus 1pt}}
\def\qline{{\vskip 3pt plus 1pt}} \def\hfline{{\skiphafline}}
\def\newpar{{\hfline\noindent}}
\def\sslash{{\slash\hskip -5pt}}
\def\slasha{{\sslash a}}
\def\mn{{\medskip\par\noindent}}
\def\bn{{\bigskip\par\noindent}}
\def\sn{{\smallskip\par\noindent}}
\def\sig{{\sigma}}
\def\olig{{oligonucleotide}}  \def\oligs{{\olig s}}
\def\oligm{{oligomer}}  \def\oligms{{\oligm s}}
\def\kmer{{$k$-mer}} \def\kmers{{$k$-mers}}
\def\dist{{distribution}}  \def\dists{{\dist s}}
\def\kdist{{$k$-distribution}}  \def\kdists{{\kdist s}}
\def\lroot{{L_{r}}}
\def\std{{std}}  \def\stds{{stds}}
\def\olig{{oligonucleotide}}  \def\oligs{{\olig s}}
\def\oligm{{oligomer}}  \def\oligms{{\oligm s}}
\def\kmer{{$k$-mer}}          \def\kmers{{\kmer s}}
\def\sigs{{$\sigma_S$}}   \def\sigl{{$\sigma_L$}}  \def\sig{{$\sigma$}} 
\def\ets{{$\eta_S$}}   \def\etl{{$\eta_L$}} 
\def\dist{{distribution}}  \def\dists{{\dist s}}
\def\foc{{occurrence frequency}}
\def\rmd{{root-mean-deviation}}
\def\bn{{\bigskip\noindent}}
\def\mn{{\medskip\noindent}}
\def\sn{{\smallskip\noindent}}
\def\mean#1{{\bar{#1}}}
\def\std{{std}}  \def\stds{{stds}}
\def\mset{{$m$-set}} \def\msets{{$m$-sets}}  
\def\MM{{\cal M}}  \def\MMs{{\MM_{\sigma}}} \def\MMi{{\MM_R}}
\def\SS{{\cal S}} \def\QQ{{\cal Q}}
\def\ie{{\it i.e.}}
\def\etal{{\it et al.}}
\def\mrep{{$m$-replica}}   \def\mreps{{$m$-replicas}}   
\def\mmul{{$m$-multiple}}  \def\mmuls{{$m$-multiples}}
\def\qr{{quasireplica}}    \def\qrs{{quasireplicas}}
\def\qrn{{quasireplication}}
%\twocolumn[\hsize\textwidth\columnwidth\hsize\csname %
%@twocolumnfalse\endcsname
\phantom{xx}
\vskip -1cm

\medskip

\title
{\bf Quasireplicas and 
universal lengths of microbial genomes}

\author{Li-Ching Hsieh$^1$, Chang-Heng Chang$^1$, 
Liaofu Luo$^3$, Fengmin Ji$^4$ and Hoong-Chien Lee$^{1,2,5,6}$}
\affiliation{
$^1$Department of Physics and $^2$Department of Life Sciences,  
National Central University, Chungli, Taiwan 320\\
$^3$Department of Physics, University of Inner Mongolia
Hohhot 010021, China\\
$^4$Department of Physics, Northern JiaoTong University,
Beijing 100044, China\\
$^5$Center for Complex Systems, 
National Central University, Chungli, Taiwan 320\\
$^6$National Center for Theoretical Sciences, Shinchu, Taiwan 300}

\date{\today}%Submitted June 10, 2003}

\begin{abstract}
Statistical analysis of distributions of occurrence frequencies of short
words in 108 microbial complete genomes reveals
the existence of a set of universal "root-sequence lengths" shared by
all microbial genomes.  These 
lengths and their universality give powerful clues to the
way microbial genomes are grown.  We show that the observed genomic
properties are explained by a model for genome growth in which
primitive genomes grew mainly by maximally stochastic duplications of 
short segments from an initial length of about 200 nucleotides (nt) 
to a length of about one million nt typical of
microbial genomes.  The relevance of the result of this study to the
nature of simultaneous random growth and information acquisition by
genomes, to the so-called RNA world in which life evolved before the
rise of proteins and enzymes and to several other topics are
discussed.
\end{abstract}

%]
\maketitle

Genomes are books of life for organisms and necessarily carry huge
amounts of information.  By and large bigger genomes carry more
information than smaller ones (there are noted exceptions).  Yet as
far as we know genomes grew and evolved stochastically, modulated by
natural selection \cite{Dawkins}.  This raises a puzzling question: 
how does genomes grow stochastically and simultaneously accumulate 
information?   This paper uses the set of all 108 sequenced complete
microbial genomes as data in exploring ideas on randomness,
entropy, information and growth with the aim of finding an answer.  
What emerges is the discovery of a set of universal
``root-sequence lengths'' governed by a simple exponential 
relation shared by all the microbial genomes and a model for 
genome growth that reproduces observed genomic 
data and that serves as at least a partial answer:    
genomes are ``\qrs''grown mainly by maximally stochastic short 
segmental duplications from random root-sequences of a universal length of 
about 200 nt.   

In what follows we do some groundwork by first discussing the relation
between the relative spectral width of a distribution of occurrence
frequencies for a set of random events and the size of the set, and
the same for the set where all frequencies are multiplied by a factor,
and then showing a simple relation between the spectral width and the
Shannon information for such sets.  We then examine certain statistical
properties of complete microbial genomes, present the results and
analyze them by way of our growth model.

%\mn
%{\bf Size, randomness and $m$-multiples}.
Consider a set of occurrence frequencies for $\tau$ types of
events, $\{f_i|\sum_{i=1}^{\tau}$$f_i$=$N\}\equiv\{f_i|N\}$, with mean
frequency $\mean{f}$ and standard deviation (\std) $\Delta$=$\langle
{(f-\mean{f})^2\rangle}^{1/2}$.  If each frequency is increased by a
factor of $m$ then $\mean{f}$ and \std\ for the new set $\{f_i' =
mf_i|mN\}$, an {\it $m$-multiple} of $\{f_i|N\}$, will both increase
by a factor of $m$ so that the relative spectral width, $\sigma\equiv
\Delta/\mean{f}$, will not change.  

%\sn 
Suppose $\{f_i|N\}$ is a set of frequencies of random events of equal
likelihood.  Then the event probability versus frequency will be
nearly a Poisson \dist\ (provided $N$$>>$$\tau$) and \std\ $\Delta_{ran}$=
$(b\mean{f})^{1/2}$, where $b$$<$1 is a factor depending only
on $\tau$ and approaches unity when $\tau$ is large.  That 
$\sigma$ scales as $N^{-1/2}$  for large $N$ is the basis for a
well known effect in
thermodynamics: the average of some measure of a random system gains
sharpness as the system gains size, and achieves infinite sharpness in
the thermodynamical (large $N$) limit.

%\sn 
Let $\{f_i|N\}$ be the frequency set for a ``small'' system $\SS_S$ of
random events, $\{f_i''|mN\}$ be the set for a ``large''
system $\SS_L$ of random events, and $\{f_i' = mf_i|mN\}$, the 
$m$-multiple of $\SS_S$, be the set for the system $\cal S'$.  
By definition both $\SS_S$ and $\SS_L$ are totally random
while $\cal S'$ is only partially so.  We have $m \mean{f} = \mean{f'}
= \mean{f''}$ and $\sigma = \sigma' = m^{1/2} \sigma''$; $\cal S'$ has
the large size of $\SS_L$ {\it and} the large $\sigma$ of $\SS_S$.
Compared to $\SS_S$, $\cal S'$ has the randomness of $\SS_S$, but
repeated $m$ times.  Compared to $\SS_L$, $\cal S'$ is less random and
more ordered by possessing a periodicity.

%\mn
%{\bf Shannon entropy and information}.  
Shannon expressed the information in a system in terms of decrease in 
uncertainty \cite{Shannon}.  
Shannon's uncertainty, or entropy, for the system 
$\{f_i|N\}$ is $H = -\sum_i (f_i/N) \log (f_i/N)$.     
We define the {\it information} of the system as 
$R \equiv H_{max} - H$, which 
accords with the thermodynamical notion that an increase in 
entropy results in a decrease in information.   We are interested in 
cases when most of the $f_i$'s are non-zero, then $H$ 
acquires its maximum value $\log\tau$ when all $f_i=\mean{f}$ and
one finds for a bell-shaped distribution 
\begin{equation}
R \equiv \log\tau - H = 
C \sigma^2 + {\cal O}\left(\sigma^3\right)\approx C\sigma^2  
\label{e:Sh_information}
\end{equation}
where the proportional coefficient $C$ is about 0.5; it is exactly 0.5 
if $\{f_i|N\}$ has a Gaussian distribution.  This reveals the
close relation between information and relative spectral width.  For
the three systems $\SS_S$, $\SS'$ and $\SS_L$ discussed above,
$R(\SS_S) = R(\SS') \approx C \sigma^2 \approx m R(\SS_L)$.  That is,
the partially ordered system $\SS'$ has the size of the larger system
$\SS_L$ and the higher information content of smaller system $\SS_S$.
Summarizing, we have: {\it (i) Given two totally random systems the
larger system carries less information; (ii) Given two systems of the
same size the one with the larger $\sigma$ has more order and carries
more information.}  We should note that since they are simply
``periodic'', \mmuls\ do not represent the best route by which a
system grows and acquires information.  In general, a large $R$ or
$\sigma$ is a necessary but not sufficient condition for high
information content.  In what follows, pending a verification of
Eq.~(\ref{e:Sh_information}), we shall use the terms information and
relative spectral width interchangeably.

%\mn
%{\bf Nucleotide sequences, \mreps\ and root-sequence length}. 
Consider now single strands of DNA, or nucleotide sequences and
view them as linear texts written in the four bases, or letters, A, C,
G, T \cite{Mantegna94,Karlin95}.  For a sequence of $L$ nucleotides
(nt) we denote by $\SS_k$, or a \kdist, the set of frequencies
$\{f_i|L\}_k$, where $f_i$ is the \foc\ of the $i^{th}$ $k$-letter
word, or \kmer, that may be obtained by moving a sliding window of
width $k$ across the genome; $\tau$=$4^k$ and $\mean{f}$=$4^{-k}L$.
To measure the information of a real genomic sequence relative to
those expected of a random sequence of the same length (and base
composition, see below) we define ``relative spectral information''
for bell-shaped distributions to be:
\begin{equation}
\MMs \equiv \sigma^2 \mean{f}/b \approx (\sigma/\sigma_{ran})^2
%\MMi \equiv R/\sigma_{ran}^2 = R/\mean{f}/b, 
\label{e:relative_sig}
\end{equation}
where $b$ is the factor associated with $ \sigma_{ran}$. 
A random sequence is expected to have $\MMs$$\approx$1. 

%\sn
Suppose $\QQ$, an {\it \mrep}, is obtained by replicating $m$ times 
a sequence $\QQ'$, then every \kdist\ $\SS_k$ of $\QQ$ is an \mmul\ of
the corresponding \kdist\ of $\QQ'$.  In particular, if $\QQ'$ is a
random sequence, then we expect (when $m$$<<$$L$) 
$\MMs(\SS_k)= m$ to a high degree of accuracy, independent of
$k$.  This motivates the following: Given sequence $\QQ$ of length
$L$ and \kdist\ $\SS_k$,  $\lroot\equiv L/\MMs(\SS_k)$ is defined as 
the {\it root-sequence length} of $\QQ$ for \kmers.  For as far
as \kmers\ are concerned, $\QQ$ (not necessarily an \mrep) has the
same relative spectral information as that of a random ``root-sequence'' of
length $\lroot$.  Only an \mrep\ of a random sequence is 
expected to have  $\lroot$'s independent of $k$ for $k<\log L/\log 4$. 

%\mn
%{\bf Complete microbial genomes}.   
The 108 complete 
microbial genomes currently in the GenBank \cite{GenBank} are
heterogeneous in length - 0.4 to 7 million bases (Mb) -
and base composition - 20 to 80\% A+T.  In most cases the
numbers of A's and T's (and of C's and G's) in a genome are 
very similar.   We therefore characterize the base composition
of a genomes by a single parameter $p$, the combined probability of A
and T, or C and G, whichever is greater.

\begin{figure}[tbh] 
\begin{center}
\includegraphics[width=3.0in,height=2.0in]
{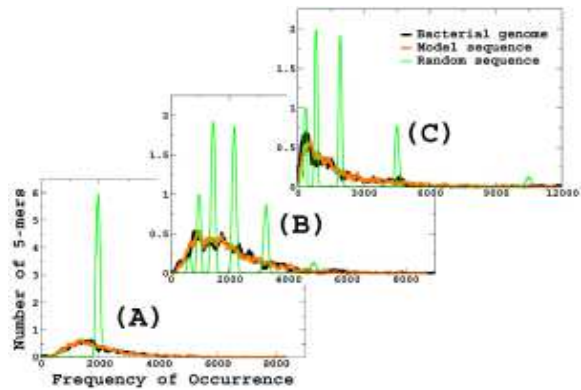}
\end{center}
\vspace{-15pt}
\caption{\label{f:ran_seqs} \small{
Comparison of 5-\dists\ of genome (black), random (green) and model 
(orange) sequences with $p$=0.5 (A), 0.6 (B) and 0.7 (C), respectively. 
The genomes are \aful\ (A), \spne\ (B) and \cace (C).}}
\vspace{-5pt}
\end{figure}

%\sn 
In Fig.~\ref{f:ran_seqs} the 5-\dists\ (green/gray curves) of random
sequences 2 Mnt long with $p$ equal to 0.5 (A), 0.6 (B) and 0.7 (C), 
respectively, are shown together with the per 2 Mnt 5-\dists\ (black) 
of three genomes with matching $p$ values: \aful\ \cite{aful} (A), 
\spne\ \cite{spne} (B) and \cace\ \cite{cace} (C). % \cite{GenBank}.  
Only the \dists\ in (A) for the 
$p=0.5$ sequences satisfy the bell-shape requirement to
make Eq.(\ref{e:Sh_information}) directly applicable.  In this case the
random \dist\ is indeed Poissonian and the genomic \dist\ is much wider, 
$\sigma_{\afuls}/ \sigma_{ran}\sim 20$ so that information-wise, 
as far as 5-mers are concerned, a 2 Mnt stretch of the \aful\ genome 
is like the 400-replica of a random sequence merely 5 knt long.

For random sequences with $p>0.5$ 
the single Poisson \dist\ is split into 
$k+1$ smaller Poisson \dists\ (Fig.~\ref{f:ran_seqs} (B) and (C)), 
one for each of the subsets of 
\kmers\ with $m$ AT's (called $m$-sets), whose respective means 
are ${\bar f}_m(p)=\mean{f} 2^k p^m (1-p)^{k-m}$, $m$= 0 to $k$. 
For the genomes the \dists\ for the $m$-sets are 
so broadened that no individual peak is discernable. 
Notably in cases when $p> 0.5$,  
the $\sigma$ for the whole \dist\ is mostly 
determined by the spread of the ${\bar f}_m$'s, which gives 
$\sigma\approx[ 2^k\left(p^2 + (1-p)^2\right)^k -1]^{1/2}$, rather 
than by the information of the sequence.  
We thus generalize 
the definition for $\MMs$ given in Eq.(\ref{e:relative_sig}) 
to be the weighted average 
over the relative spectral information of the \msets:
\begin{equation}
\MMs \equiv  
\sum\nolimits_{m=0}^k L^{-1}\left(2^k (k,m){\bar f}_m\right) 
\sigma^2_{k,m}\mean{f}_m/b
\label{e:relative_sig_mean}
\end{equation}
where $(k,m)$ is a binomial, $\sum_m 2^k (k,m) {\bar f}_m=L$.  
Since A and T (and C and G) are counted together and 
the number of each monomer in the sequence is fixed, 
the binomial factor is taken to be $b=1-2^{1-k}$. 
%These definitions can now be applied to a sequence with any 
%$0.5\ge p\ge 1$ and they are used in our computation. 
In practice, to circumvent large fluctuations in $\sigma_{k,m}$   
induced by small unevenness in the A/T (or C/G) contents 
- this can occur when ${\bar f}_m$ is very large at $k$=2 and 3 -
each frequency is divided by a factor 
$(2^k/p^m(1-p)^{k-m})\prod_s p_s^{m_s}$, where $s$ runs over the four 
bases and $\sum_s m_s=k$.  
To verify Eq.~(\ref{e:Sh_information}), we also define a 
``relative Shannon information'' $\MMi$ where $\sigma^2_{k,m}$ in 
Eq.~(\ref{e:relative_sig_mean}) is replaced by $R_{k,m}$.

\begin{figure} [tbh] 
\begin{center}
\includegraphics[width=2.9in,height=2.5in]%[width=8cm,height=7.0cm]
{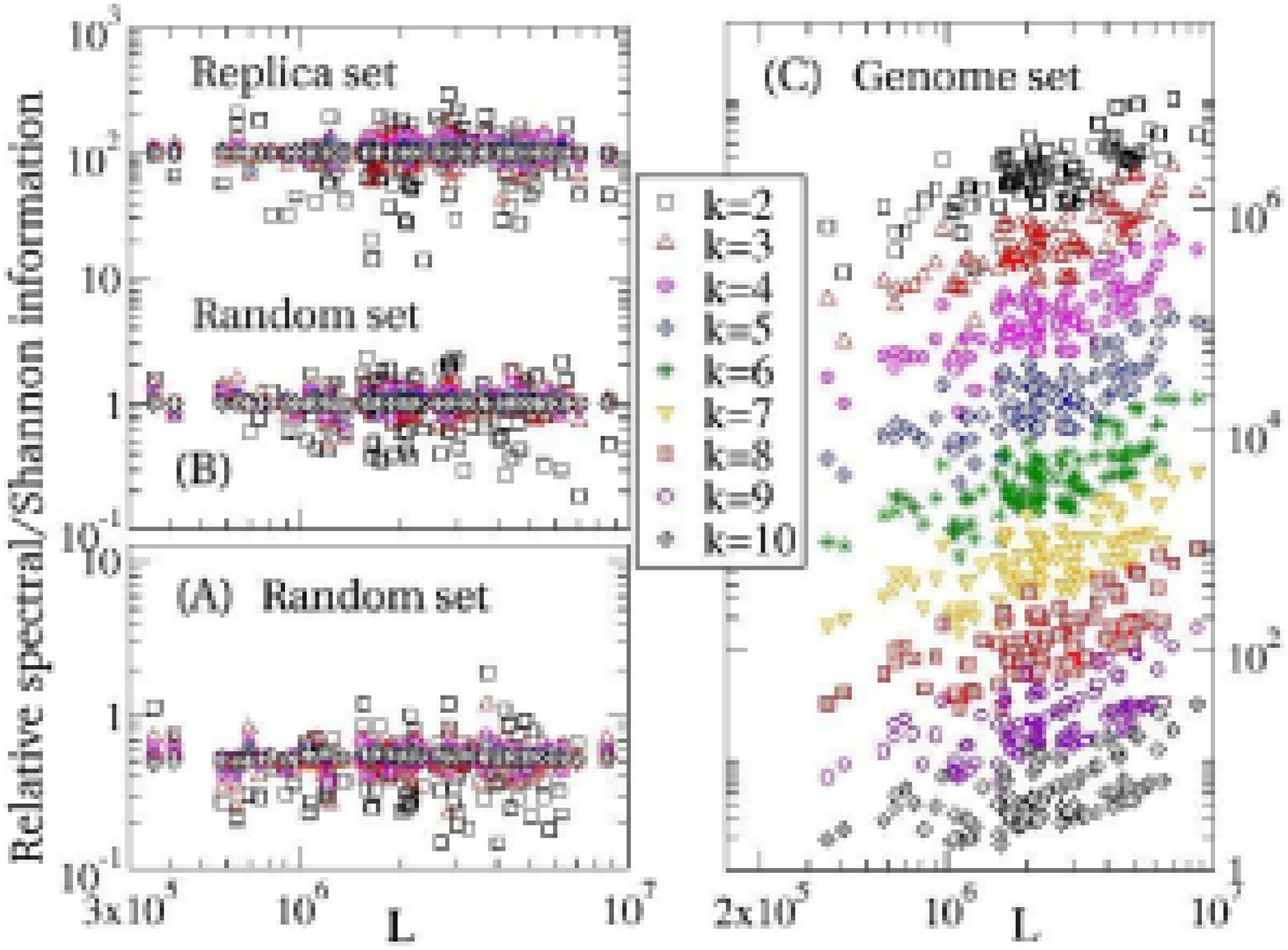}
\end{center}
\vskip -15pt
\caption{\label{f:m_vs_k} \small{
The relative Shannon information ($\MMi$) and relative spectral 
information ($\MMs$) for three sets of sequences.   
(A) $\MMi$ for the random set; (B) $\MMs$ for the 
random and replicas sets; (C) $\MMs$ for the genome set.}} 
\vskip -5pt
\end{figure}

%\mn
%{\bf Universal root-sequence lengths of microbial genomes}.  
Fig.~\ref{f:m_vs_k} shows log-log plots of $\MMs$ and $\MMi$ versus
sequence length $L$ computed from a ``genome'' set composed of 108
complete microbial genomes \cite{GenBank} and two control sets with
lengths and base compositions matching those in the genome set: a
``random'' set of random sequences and a ``replica'' set of
100-replicas of random sequences.  The results for the control sets
are shown in the left-panel ((A) and (B)) of Fig.~\ref{f:m_vs_k} and
they are essentially independent of $k$, sequence length $L$ and base
composition $p$ and have the expected values: (A) $\MMi= C=
0.514$$\pm$0.062 - this verifies Eq~(\ref{e:Sh_information}); (B)
$\MMs=1.02$$\pm$0.11 for the random set and $\MMs= 101$$\pm$12 for the
replica set.  Each set contains 972 pieces of data and in each plot
about 50\% of the error comes from data for $k$=2 (``$\Box$'' in the
figure) and 25\% from $k$=3 (``$\triangle$'').  This is because
$\mean{f}_{k,m}$ for these cases are very large and magnify
fluctuations in $\sigma_{k,m}$ in Eq~(\ref{e:relative_sig_mean}).

\begin{figure} [tbh]
\begin{center}
\includegraphics[width=6.5cm,height=4.0cm]
{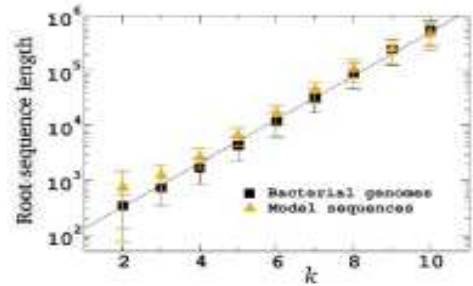}
\end{center}
\vskip -0.5 cm
\caption{\label{f:Lr_vs_k} \small{
$\lroot$ versus $k$ extracted from relative spectral information 
of the genome set (squares) and a set of 108 model sequences (triangles) 
whose lengths and base compositions match those in the genome set.
Line shows mean of Eq.~(\protect\ref{e:exponential}).}}
\vskip -0.3 cm
\end{figure}

The right-panel of Fig.~\ref{f:m_vs_k} shows $\MMs$ for the genome
set, where each piece of data was multiplied by a factor of $2^{10-k}$
to delineate data into different $k$ groups for better viewing.  Still
essentially $p$-independent, the data are otherwise entirely different
from those of the control sets: (i) For given $k$ they form a band
(std is about 50\% of mean) that depends linearly on $L$, implying
that $\lroot= L/\langle\MMi\rangle$ is a {\it universal}
root-sequence length, \ie, the same for all microbial genomes. (ii)
For given $L$, the mean $\log\MMi$ decreases approximately linearly
with increasing $k$, such that the universal lengths (squares in
Fig.~\ref{f:Lr_vs_k}) satisfy an approximate exponential relation of
the form
\begin{equation}
\lroot(k) = \Lambda\times t^k; \quad 2\le k \le 10
\label{e:exponential}
\end{equation}
with $\Lambda=48\pm24$ nt and $t=2.53$.  
%The relation  $\lroot(2)$$\sim$200 nt implies that for 2-mers a 1 Mnt genome 
%is equivalent to a 5000-replica of a 200 nt random sequence, whereas 
%$\lroot(10)$$\sim$360 knt states that for 10-mers a 1 Mnt genome much more 
%closely approximates a 1 Mnt random sequence.  
We remark that the property revealed by the log-log plot 
in Fig.~\ref{f:m_vs_k} (C) is different from Zipf's law
\cite{Zipf} and its variants that have been much discussed for \kmer\
frequencies ($k\ge$ 6) in biological sequences \cite{Mantegna94,Luscombe02}. 

%\mn
%{\bf Model for genome growth - quasireplication}.
The universality of the $\lroot$'s suggests the existence
of a universal mechanism for (microbial) genome growth from
proto-genomes of a universal initial length.  The very large values 
of $\MMs$ (hence the shortness of the $\lroot$'s) for the smaller 
$k$'s imply a mechanism involving a great deal of 
replication or duplication.  One obvious mechanism, growth mainly by
whole-genome replications \cite{Ohno70,Hughes01} is
ruled out because that would yield $k$-independent $\lroot$'s, 
contrary to data.  The observed strong $k$-dependence of
$\lroot$ suggests a more complex duplication process.  

We show that ``universal genomes'' generated in a simple and
biologically plausible growth model \cite{Hsieh02,Hsieh03} possess
properties similar to those of microbial genomes.  In the model the 
initial condition of a genome is a random sequence about $L_0\approx$ 200 nt 
long with a base composition characterized by a given value of $p$.  
The condition $L_0<\lroot(2)\approx$ 300 nt is
necessary if the large values of $\MMs$ for the small $k$'s are to be
attained.  The genomes then grows by random short
segmental duplications - or {\it \qrn} - possibly modulated by random
single mutations.  The model shares some features with those used to
explain the power-law behavior of the \foc\ of genes in genomes
\cite{Yanai00,Qian01}, except that there the units of duplication are
genes, not the short \oligs\ used here.  The \qrn\ process is
maximally stochastic: a segment of length $l$, chosen according to the
probability density function $f(l)=1/(a n!)  (l/a)^n e^{-l/a}$, is
copied from one site and inserted into another site, both randomly
selected.  The Erlang function, the integer $n$=2 and the length scale
$a$=6.7 (nt) were determined by data, implying a typical length of
$20\pm 12$ nt for the duplicated segments.

In Fig.~\ref{f:Lr_vs_k} the $\lroot$'s (triangles) extracted from a
set of 108 model sequences with length and base composition matching
those in the genome set and generated {\it in silico} by \qrn\ are
compared with the $\lroot$'s for the genome set (squares).  The two
sets of lengths essentially agree although those from the model
sequences have a slightly weaker $k$-dependence.  The \kdists\ of
5-mers computed from three representative model sequences with $p$=
0.5, 0.6 and 0.7, respectively, are shown as orange curves in
Fig.~\ref{f:ran_seqs}.  These results show that our very simple growth
model, in being able to produce sequences that faithfully exhibit
signature global properties of microbial genomes, likely has captured
the essence of the real genomic growth process.  We believe that without  
the two main ingredients of our model, very short initial genome length 
($\lesssim$ 200 nt) and random duplication of short segments, no simple 
growth model can produce the results shown in Fig.~\ref{f:Lr_vs_k}. 

%\mn
%{\bf Microbial genomes are quasireplicas}.
We call sequences generated in our model {\it quasireplicas}.  Unlike
an \mrep, a \qr\ is globally aperiodic.  If the length of a \qr\ is
$L$, then the maximum $k$ for which Eq.~(\ref{e:exponential}) applies
is given by $k_{max}\approx\log L/\log 4$.  For $k\le k_{max}$ a \qr\
acts as an \mrep\ where $m=L/\lroot(k)$, while for $k>>k_{max}$ it
appears essentially as a random sequence.  Quasireplicas are partially
ordered, highly complex and evidently capable of carrying large
amounts of information.  Our study shows that microbial genomes are
self-organized quasireplicas belonging to the class given by
$\Lambda\approx 48$ nt and $t\approx 2.53$ of
Eq.~(\ref{e:exponential}).  It is a common feature of complex
self-organized systems to exhibit power-law relations \cite{Bak} and
the relation between our model and Eq.~(\ref{e:exponential}) is being
explored.

Quasireplicas are extremely robust.  We have verified that, provided
the typical duplicated segment length is significantly greater than
$k_{max}$, quasireplication (including simple replication) of a \qr\
begets a longer \qr\ of the same class.  We are currently studying
eukaryotic genomes in terms of \qrs\ and results will be reported
elsewhere.  Based on findings obtained so far we make the following
proposition: the ancestors of microbial genomes underwent a
fundamental transition in their growth and evolution shortly after
they had reached a length of about 200 nt, by which time they had
acquired a rudimentary duplication machinery, and then grew (and
diverged) mainly by short-segment \qrn\ to become quasireplicas of the
order of 1 Mnt long.  Assuming this proposition to be substantially
true we mention some implications.

%\mn
%{\bf A new major transition and the RNA world}.
It seems that by adopting the natural method of \qrn\ for growth, 
microbial genomes also adopted a superb strategy for information 
acquisition and accumulation, and in doing so they left a clear 
evolutionary track: the universal root-sequence
lengths.  This seems to put the onset of \qrn\ in the position of
being another, hitherto unknown, major transition in evolution
\cite{Smith}.  Before that transition the ancestral genomes and their
prebiotic precursors somehow acquired a store of
information - including a rudimentary duplication machinery - and
after the transition the genomes evolved via 
\qrn.  The smallness of the genomes at the transition - less than a
quarter of the size of a present-day gene coding for a typical enzyme
- implies that at the time they must have lived in an ``RNA world''
devoid of proteins \cite{Gilbert86,Joyce02}.  It is likely that the
ribozymes that made up the duplication machinery at that time were not
much bigger than the smallest ribozymes now extant, about 31 to 50 nt
\cite{Forster87}; hence the average duplicated segment length of about
20 nt would have been effective in propagating coded information which,
presumably, could later be locally varied under the combined forces of
mutation and natural selection for adaptation to new purposes.  An RNA
world reigned no more than 600 million years, from about 4.2 (when the
earth cooled down) to 3.6 (when protein must have appeared) billion
years ago - probably even much shorter \cite{Joyce02}.  It is not
necessary that genomes grew to their respective current lengths during
that period.  It is sufficient that during that period, growth by
short segmental \qrn\ produced basic quasireplicas which, after the
rise of proteins and enzymes, could be further expanded upon via
quasireplication by duplicating longer segments, including genes
\cite{Yanai00,Qian01}.  Whatever path the genomes actually took, their
rate of evolution must have been tremendous during the RNA era, and
\qrn\ probably had a better chance of meeting that challenge than any
other alternative.

This work is supported in part by the grant 91-2119-M-008-012 from 
the National Science Council, ROC.  HCL thanks H.Y. Lee and Ceaga Lee 
for discussions.

\end{document}